\documentclass[9pt,twocolumn,twoside]{opticajnl}
\journal{opticajournal} % use for journal or Optica Open submissions

\setboolean{shortarticle}{false}
\usepackage{lineno}
\usepackage{graphicx}% Include figure files
\usepackage{dcolumn}% Align table columns on decimal point
\usepackage{amsmath,amssymb}% bold math
\usepackage{svg} % import svg images
\usepackage{lipsum}
\usepackage[utf8]{inputenc}
\usepackage[T1]{fontenc}
\usepackage{etoolbox}
\usepackage{color}
\usepackage{soul}
\usepackage{siunitx}
\usepackage[version=3]{mhchem}
\usepackage[acronym,shortcuts]{glossaries}
\usepackage{url}
\title{Optical pumping enhancement of a free-induction-decay magnetometer}

\author[1,*]{Dominic Hunter}
\author[1]{Marcin S. Mrozowski}
\author[1]{Allan McWilliam}
\author[1]{Stuart J. Ingleby}
\author[1]{Terry E. Dyer}
\author[1]{Paul F. Griffin}
\author[1]{Erling Riis}

\affil[1]{Department of Physics, SUPA, University of Strathclyde, Glasgow G4 0NG, United Kingdom}

\affil[*]{d.hunter@strath.ac.uk}

\begin{abstract}

Spin preparation prior to a free-induction-decay (FID) measurement can be adversely affected by transverse bias fields, particularly in the geophysical field range. A strategy that enhances the spin polarization accumulated before readout is demonstrated, by synchronizing optical pumping with a magnetic field pulse that supersedes any transverse fields by over two order of magnitude. The pulsed magnetic field is generated along the optical pumping axis using a compact electromagnetic coil pair encompassing a micro-electromechanical systems (MEMS) vapor cell. The coils also resistively heat the cesium (Cs) vapor to the optimal atomic density without spurious magnetic field contributions as they are rapidly demagnetized to approximately zero field during spin readout. The demagnetization process is analyzed electronically, and directly with a FID measurement, to confirm that the residual magnetic field is minimal during detection. The sensitivity performance of this technique is compared to existing optical pumping modalities across a wide magnetic field range. A noise floor sensitivity of $\mathbf{238}\,$fT$/\boldsymbol{\sqrt{\mathrm{Hz}}}$ was achieved in a field of approximately $\mathbf{50}\,\boldsymbol{\mu}$T, in close agreement with the Cram{\'e}r-Rao lower bound (CRLB) predicted noise density of $\mathbf{258}\,$fT$/\boldsymbol{\sqrt{\mathrm{Hz}}}$. 

\end{abstract}

\setboolean{displaycopyright}{false} % Do not include copyright or licensing information in submission.

\begin{document}

\maketitle

\section{Introduction}
\vspace{-0.3cm}
Extensive efforts have been devoted toward developing optically pumped magnetometers (OPMs) that operate at zero field \cite{cohen1970diverses, castagna2011measurement, jimenez2014optically}, exploiting the well-established spin-exchange relaxation-free (SERF) mechanism \cite{allred2002high, shah2009spin, li2018serf}. Such devices can achieve exceptionally high sensitivity \cite{dang2010ultrahigh}, and are therefore well-suited to applications demanding the capability to resolve fT-level signals, e.g., magnetoencephalography (MEG) \cite{boto2017new}. Sensors operating in the SERF regime are already commercially available at sensitivities below $10\,\mathrm{fT/\sqrt{Hz}}$ with a bandwidth of $135\,$Hz \cite{osborne2018fully}. Accordingly, they offer an attractive alternative to superconducting quantum interference devices (SQUIDs), particularly in MEG applications, as these compact and flexible devices can be easily integrated into custom mounting hardware \cite{boto2018moving}. However, the narrow magnetic resonances essential for high sensitivity operation in SERF devices also imposes limitations on both sensor bandwidth and dynamic range. This restricts the implementation of these sensors to low-field environments that are often conditioned using both passive and active field compensation techniques \cite{holmes2019balanced}.  \\
\indent The exceptional sensitivity achievable with SERF operation is attributable to suppression of spin-exchange collisions that occur between alkali atoms when operating close to zero field with dense atomic ensembles. Several studies have been conducted that extend the utility of spin-exchange suppression, enabling $\mathrm{fT/\sqrt{Hz}}$ sensitivity operation in bias field's of several $\mu$T \cite{gerginov2017pulsed, guo2019compact, lucivero2022femtotesla}. Such devices provide a framework for unshielded sensing that could become a valuable resource in many research areas including geophysical \cite{beggan2018observation}, space science \cite{korth2016miniature}, GPS-denied navigation \cite{canciani2021magnetic}, and biomedical applications \cite{limes2020portable}. These sensors rely on the light-narrowing phenomenon \cite{appelt1999light, scholtes2011light}, which requires close to unity spin polarizations to be generated. Therefore, maintaining sufficient optical pumping efficiency is crucial in both maximizing signal-to-noise ratio (SNR) and lowering spin-exchange contributions to the transverse relaxation rate, $\gamma_{2}$. \\
\indent This study employs an OPM based on the free-induction-decay (FID) measurement protocol \cite{grujic2015sensitive, hunter2018waveform, hunter2022accurate}. This modality exhibits a wide and tunable bandwidth given the flexibility in digital signal processing (DSP) that can be used to analyse the FID data \cite{hunter2022accurate, wilson2020wide}. A notable benefit of FID-based sensors is their accuracy, as optical pumping and detection are separated temporally. Consequently, this allows the polarized spins to precess freely at the Larmor frequency without being perturbed by intense pumping light, thereby considerably lowering light-shift systematics compared to continuous-wave (cw) pumping schemes \cite{ingleby2022digital}. Moreover, these sensors are robust as they are commonly operated in a free-running mode and enable direct Larmor frequency extraction. Additionally, spin-exchange suppression can be exploited to extend the sensor dynamic range to bias fields exceeding the Earth's field through efficient spin preparation. \\
\indent The presence of strong magnetic fields can hinder the optical pumping dynamics as the atoms experience a torque that deflects the generated spin polarization from the beam propagation (optical pumping) axis. This issue can be circumvented by, for example, resonantly driving the atoms at the Larmor frequency by modulating either the amplitude \cite{grujic2015sensitive} or frequency \cite{hunter2018free} of the pump light. Additionally, one can null the external magnetic field contributions during the spin preparation stage \cite{gerginov2020scalar}. Although effective, both these techniques require prerequisite knowledge regarding the field of interest, and feedback to maintain optimal conditions. \\
\indent In this work, we exploit a simple and practical approach, hereby referred to as enhanced optical pumping (EOP), that facilitates efficient spin polarization buildup throughout a range of bias field conditions. This is achieved by applying a strong field, $\vec{B_{p}}$, of several mT along the quantization axis during spin preparation to negate the detrimental effects of transverse field components. Additionally, the resistive heating produced by the coils generating $\vec{B_{p}}$ is utilized to elevate the vapor cell temperature and reach an optimal atomic density. This provides an effective way of exploiting the FID sensor dead-time whilst simultaneously enhancing the optical pumping efficiency. \\
\indent The approximate optimum atomic density occurs when the ratio of $\gamma_{2}$ and SNR is maximized \cite{jimenez2009sensitivity}. However, delivering heating power to the vapor cell can often adversely affect the instrument's performance. For instance, the alkali atoms are most commonly heated by passing current through a resistive element in contact with the cell. Current noise flowing through the heating element is converted to magnetic fluctuations that can lift the sensor noise floor. Furthermore, magnetic resonance broadening and additional systematics can be induced by subsequent stray fields. Oscillating currents are often used at frequencies far exceeding the atomic bandwidth to alleviate these issues \cite{tayler2022miniature}. However, heating at MHz frequencies far beyond the Larmor precession rate is often necessary when applied to total-field OPMs \cite{ingleby2022digital}, which is typically inefficient. There have been non-electrical heating methods adopted in the past such as optical heating \cite{sheng2017microfabricated} and hot air systems \cite{kominis2003subfemtotesla}, although these techniques require a great deal of power and are often restricted to a laboratory setting. The heating strategy proposed in this work ensures no current is flowing through the heaters during spin readout, hence is immune to systematics and magnetic noise contributions that often contaminate other OPM technologies. This is made possible by demagnetization electronics that enable the current through the heating coil to be switched from $1.4\,$A to within $10\,\%$ of the MOSFET leakage current ($< 50\,$pA) over a period of approximately $2500\,$ns. 

\section{Experimental methodology}
\subsection{FID magnetometer setup}
\vspace{1pt}
\begin{figure*}
	\centering
	\includegraphics{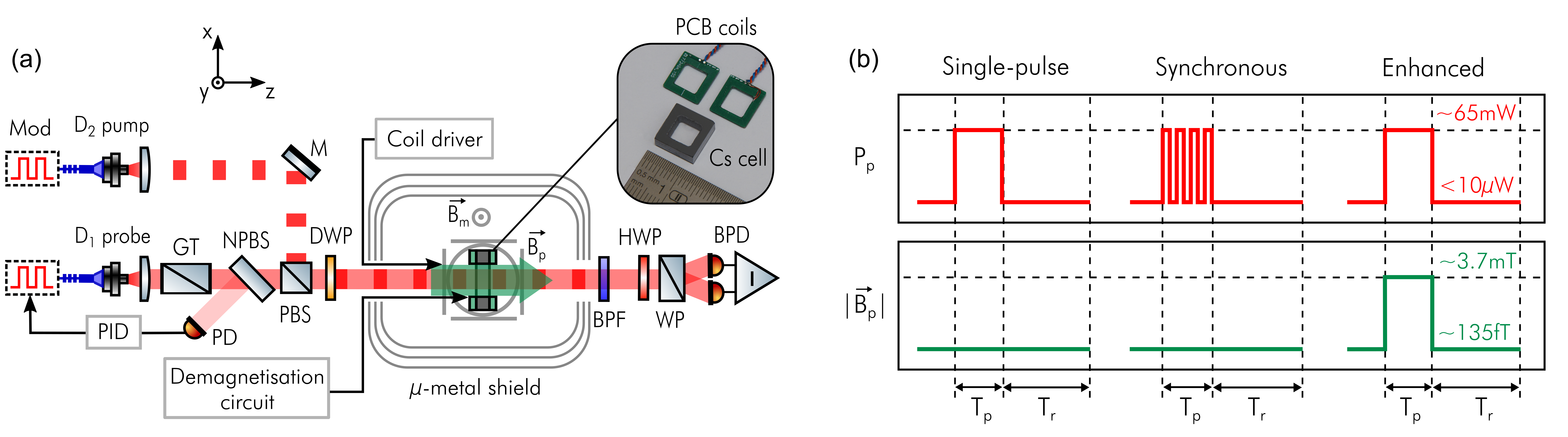}
	\caption{(a) Simplified schematic of a FID OPM utilizing a two-color pump-probe configuration: GT, Glan-Thompson polarizer; NPBS, non-polarizing beamsplitter; PD, photodiode; M, mirror; PBS, polarizing beamsplitter; DWP, dual-wavelength waveplate; HWP, half-wave plate; BPF, bandpass filter; WP, Wollaston prism; BPD, balanced photodetector; PCB, printed circuit board. (b) Depiction of various optical pumping techniques that can be employed. In the single-pulse scheme, the atoms are continuously pumped at peak optical power throughout the period, $T_{p}$. Synchronous pumping modulates the light resonantly at the Larmor frequency. The EOP scheme applies an additional longitudinal field, $\vec{B_{p}}$, of approximately 3.7\,mT along the $z$-axis, synchronized with the optical pulse. The vapor cell is positioned between two PCB coils to generate $\vec{B_{p}}$, which are also used for resistive heating. Pump light interacting with the atoms is extinguished to $<\,10\,\mu$W during the readout period, $T_r$. The PCB coils are demagnetized to within $10\,\%$ of the field produced by the MOSFET leakage current ($|\vec{B_{p}}| \sim 135\,\mathrm{fT}$) at $2.5\,\mu$s.}
	\label{Experimental setup}
\end{figure*}
A simplified schematic of the experimental arrangement is illustrated in Fig. \ref{Experimental setup}(a), showing a two-beam OPM operating in the FID regime. The sensor head consists of a $3\,$mm thick micro-electromechanical systems (MEMS) Cs vapor cell with nitrogen (N$_2$) buffer gas \cite{dyer2022micro}. As performed in previous work \cite{hunter2018free}, the collisional broadening and shift in the optical spectrum measured against a Cs reference cell determined the internal buffer gas pressure to be approximately 220 torr. \\
\indent Optical pumping and detection are performed using two co-propagating probe and pump beams tuned to the Cs $D_1$ and $D_2$ transitions, respectively. Both beam widths were set to a diameter ($1/e^2$) of $3.1\,$mm. While the $895\,$nm probe light remains linearly polarized, the $852\,$nm pump beam becomes circularly polarized after passing through a dual-wavelength multi-order waveplate. This enables optical pumping and detection to be optimized without sacrificing sensitivity, as opposed to launching the beams at a slight angle relative to one another \cite{gerginov2017pulsed}. A polarizing beamsplitter is used to combine both beams prior to traversing the waveplate and illuminating the vapor cell. Overlap between both beams within the interaction region is optimized based on the maximum FID signal amplitude.  \\
\indent A high power ($\leq 600\,$mW) single-frequency laser is used to optically pump the alkali spins into a highly polarized state. The pump light is resonant with the $F = 3 \longrightarrow F^{\prime}$ hyperfine transition of the Cs MEMS cell. The transition is collisionally broadened to a full width at half maximum (FWHM) linewidth of $3.7\,$GHz and shifted by $-1.6\,$GHz due to the N$_{2}$ buffer gas. Pumping on this transition provides efficient recycling of atoms from the $F = 3$ ground state so they can subsequently contribute to the signal \cite{schultze2015improving}. Light narrowing is exploited to partially suppress spin exchange as most of the atomic population is transferred to the $F=4$ ground state when pumped optimally, and thus cannot exchange spin due to angular momentum conservation \cite{scholtes2011light}. \\
\indent The probe laser produces a cw beam that is $20\,$GHz blue-detuned from the $F = 4 \rightarrow F^{\prime} = 3$ Cs transition. This mitigates broadening of the magnetic resonance during detection, by reducing residual optical pumping whilst maintaining an appreciable light-atom interaction strength to maximize signal amplitude. The Glan-Thompson polarizer purifies the probe beam polarization and converts polarization noise, e.g., arising from the fiber, into amplitude noise. A non-polarizing beamsplitter separates the light equally between a monitor photodiode and the vapor cell. The probe power ($\approx 450\,\mu$W) prior to the vapor cell was actively stabilized, to within $0.4\,\%$, using an analog PID controller (SRS SIM960) that adjusts the RF power supplied to an acousto-optic modulator (AOM). \\
\indent The measurement bias field, $\vec{B_{m}}$, strength and direction can be controlled by applying currents through a set of three-axis coils that encapsulate the cell. The experiments performed here used only a single transverse axis coil to produce a field with magnitude, $B_{y}$, along the $y$-axis. This was driven by a custom current source with a $\pm\,75\,$mA range and a noise level considerably below the noise floor of the sensor \cite{mrozowski2023}. Accordingly, the sensor's dynamic range can be evaluated up to $B_{y} \approx 50\,\mu$T using this coil. The whole assembly is placed inside a three-layer $\mu$-metal shield that attenuates environmental magnetic variations, and ambient fields down to nT levels. 

\subsection{Optical pumping schemes} \label{OP schemes}
\vspace{1pt}
\indent Figure \ref{Experimental setup}(b) shows two optical modulation schemes that can be employed during the spin preparation period, $T_{p}$, dedicated to optical pumping \cite{hunter2018free}. The first method uses single-pulse (SP) modulation where the maximum available light intensity interacts with the atoms throughout the pumping phase, before switching to approximately zero during the detection period, $T_{r}$. One can also optically pump the atoms by resonantly modulating the light intensity at the Larmor frequency $\omega_{L} = \gamma |\vec{B}|$, where $|\vec{B}|$ is the magnetic field magnitude and $\gamma$ is the gyromagnetic factor dependent on the atomic species, i.e., $\sim 3.5\,$Hz/nT for Cs. This technique is known as synchronous modulation \cite{grujic2015sensitive}. The peak optical power is $\sim 65\,$mW after the pump light has traversed beam-conditioning optics, an AOM, and a fiber-coupling stage. The AOM's extinction ratio ensures that $<10\,\mu$W of pump light interacts with the atoms during the off state. A spectral bandpass filter is placed in front of the balanced photodetector to attenuate $852\,$nm light whilst allowing $895\,$nm light to pass with $>90\,\%$ transmission, thus avoiding saturation and lowering optical noise contributed by the pump light. \\
\indent This work demonstrates an alternative optical pumping strategy, EOP, which is exploited by mounting the vapor cell between two compact printed circuit boards (PCBs) that have a square central aperture for optical access, as seen in Fig. \ref{Experimental setup}(a). The PCB assembly serves two purposes: to generate a strong field, $\vec{B_{p}}$, along the beam propagation axis; and maintaining an optimal atomic density within the vapor cell through resistive heating. The technique is illustrated in Fig. \ref{Experimental setup}(b) with $|\vec{B_{p}}|$ set to several mT and applied along the optical pumping ($z$-axis), synchronized with the optical pulse over the pump period. Subsequently, the PCB coils producing $\vec{B_{p}}$ are demagnetized by rapidly lowering the current flow to zero such that only $\vec{B_{m}}$ persists during spin readout. Details regarding the demagnetization circuitry are discussed in Section \ref{demag}. As the strength of $\vec{B_{p}}$ supersedes $\vec{B_{m}}$ by at least two orders of magnitude, the macroscopic spin magnetization is pinned to the $z$-axis. This negates the adverse impact of any transverse field components, including $\vec{B_{m}}$, during spin preparation. \\
\indent The copper tracks on the PCB are printed with a square spiral pattern on both sides of the two-layer PCB. A via is used to electrically connect both layers. The compact bi-planar stack can thus generate stronger magnetic fields compared to a single layer PCB, with the field-to-current ratio theoretically modelled to be $2.7\,\mu$T/mA at the center of the vapor cell. Additionally, using multiple layers enables more coil turns in a smaller footprint for increased heating efficiency. The vapor cell temperature can be controlled by either adjusting the duty cycle or peak current of the applied current pulse flowing through the PCB coils. The duty cycle also impacts the time dedicated to optically pumping the atoms using EOP as the optical and magnetic pulses are synchronized. The spin polarization was found to be close to saturation after pumping for $T_{p} \approx 88\,\mu$s. This is equivalent to a $8.8\,\%$ duty cycle for the pump-probe cycle repetition rate, $f_{d} = 1\,$kHz, which was kept consistent in the experiments performed here. A peak current of $1.4\,$A heats the vapor cell to a temperature of $\mathrm{88\,^{\circ}C}$ providing the optimal atomic density. \\
\indent $T_{p}$ was kept consistent when employing EOP or SP modulation to provide a valid comparison between both techniques. Synchronously pumping the atoms required longer to reach a steady-state spin polarization compared to EOP, with $T_{p}$ set to around $286\,\mu$s. This can be attributed to the light intensity being low for most of the pump phase as the optimal duty cycle of the square-wave modulation was approximately $30\,$\%. As a result, $T_{r}$ was shorter for synchronous operation since $f_{d}$ was kept constant in each case. Additionally, gated heating at $0.5\,$Hz was applied to the vapor cell when utilizing synchronous or SP optical pumping, with measurements conducted when no current was flowing through the PCB. This was to ensure that magnetic noise from the heater was not contributing to the noise floor of the sensor using these techniques. 

\subsection{Demagnetization electronics}
\vspace{1pt}
\label{demag}
Current flowing through the coils cannot appear or disappear instantaneously due to their inductive nature. The coil acts as an open circuit after being magnetized and subsequently switched-off. This sudden change in current causes back EMF to be generated with opposing polarity to the supply voltage until the current exponentially decays to zero. During this process, the induced voltage is high ($10\,$s of kV) which can easily damage switching electronics. A diode can be added in parallel to the coil to clamp this back EMF to the forward bias voltage of the diode, and allow for safe demagnetization. However, this approach is relatively slow as the forward bias voltage limits the maximum energy at which the coil can demagnetize. This process can be sped up drastically using a Zener diode by exploiting its avalanche breakdown mechanism to rapidly demagnetize the coil \cite{PhysRev1954avalanche}. This clamps the back EMF to the avalanche breakdown voltage of the Zener, such that more energy is dissipated at a faster rate leading to more rapid demagnetization. \\
\begin{figure}[t]
	\centering
	\includegraphics[scale = 1.2]{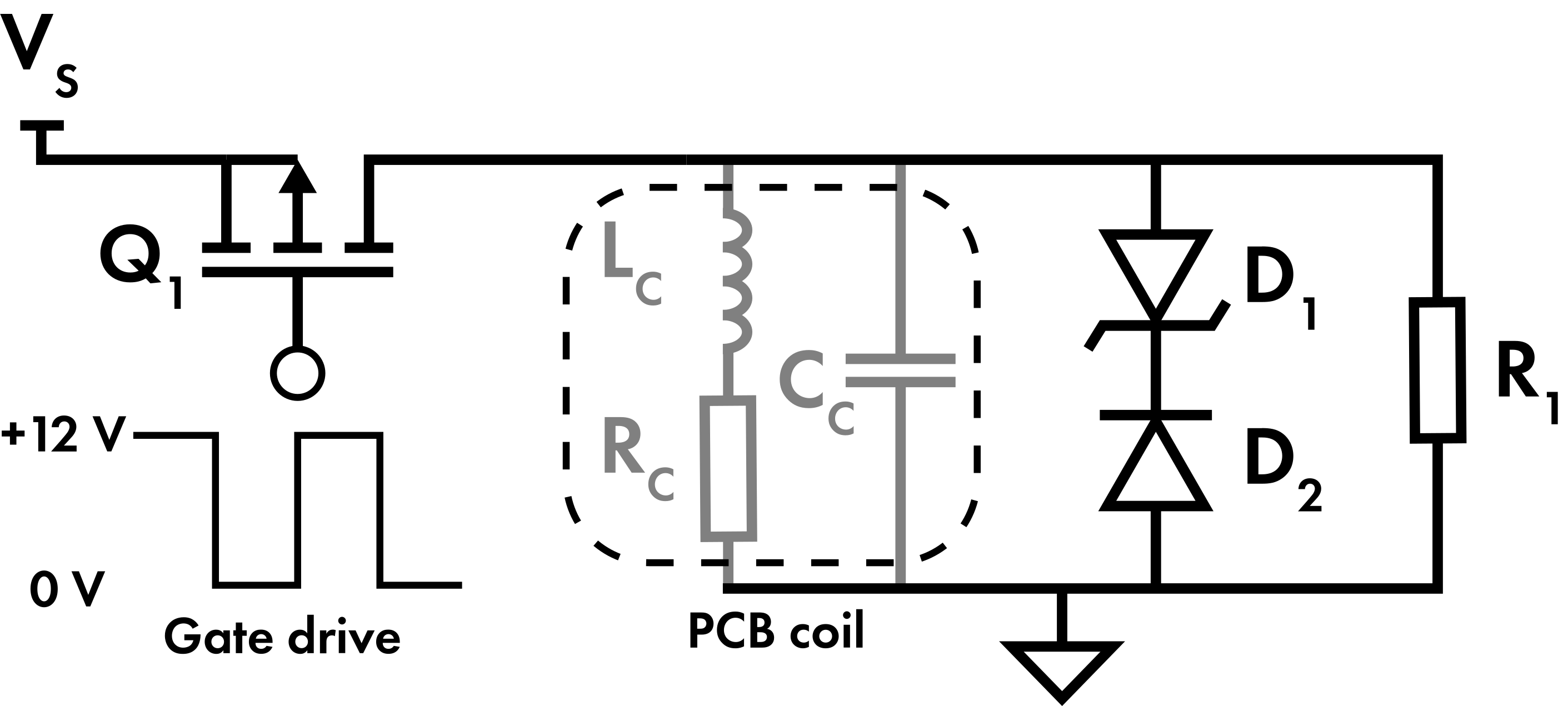}
	\caption{Simplified schematic of the demagnetization circuitry used to produce $\vec{B_{p}}$ for EOP, and to drive the resistive heating applied to the vapor cell. A pair of PCB coils (parasitics represented in light grey) were found to have an inductance $L_C \approx$\,10\,$\mu$H, interwinding capacitance $C_C \approx$\,10\,pF, and resistance $R_C \approx$\,4\,$\Omega$.}
	\label{schematic coil demag}
\end{figure}
\indent The circuit designed to rapidly magnetize and demagnetize the PCB coils is presented in Fig.~\ref{schematic coil demag}. A P-channel MOSFET ($Q_1$) is used as a switch to control the current flow through the coil. $Q_1$ is controlled with a function generator through a gate driver circuit that allows for a fast switching rate of the MOSFET, in addition to acting as a buffer. The demagnetization is managed by diodes $D_1$ and $D_2$. The transient-voltage-suppression (TVS) diode, $D_1$, works similarly to a Zener diode although enables more energy dissipation due to its larger area p-n junctions \cite{microsemi_134}. The TVS breakdown voltage was selected to be close to the maximum permitted drain voltage of $Q_1$, shortening the demagnetization time while preventing damage to $Q_1$. The fast recovery rectifying diode, $D_2$, has the sole purpose of preventing $D_1$ from conducting during the magnetization process. $R_1$ is a 100\,$\Omega$ low inductance wirewound resistor (Vishay WSN) connected in parallel with the coil, and serves as the final demagnetization device after the induced voltage falls below the $D_2$ forward voltage. $R_1$ also serves as a dampening device for the interwinding capacitance $C_c$ of the coil. At the end of the demagnetization process, the current flowing through the coil is equivalent to the transistor drain leakage. The circuit is powered from a triple output power supply. The strength of $\vec{B_{p}}$ and the heating power are controlled by adjusting supply voltage $V_S$ using the $2.5\,$A output. \\
\begin{figure}[t]
	\centering
	\includegraphics[scale = 0.51]{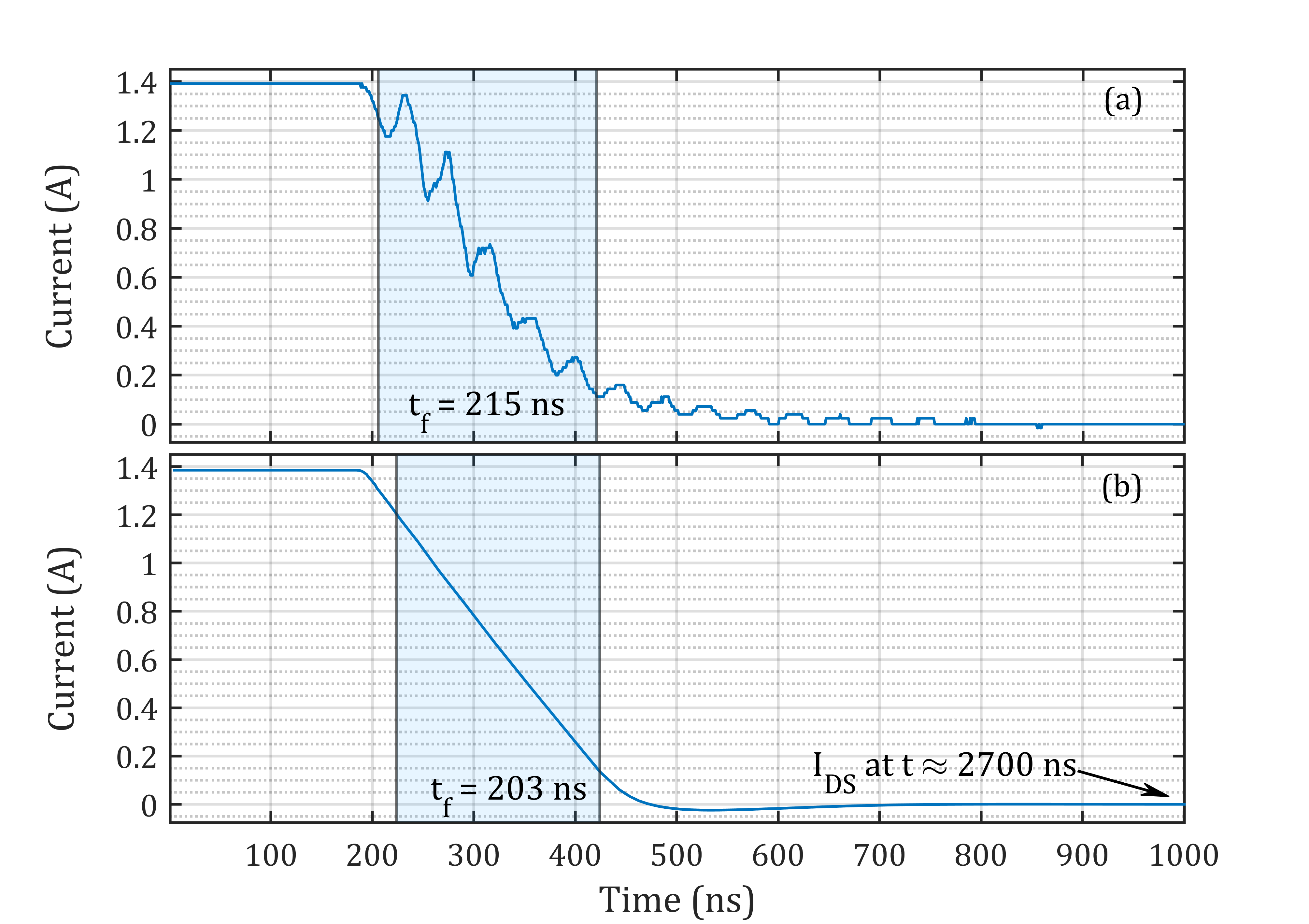}
	\caption{Transient response of the demagnetization circuit derived (a) experimentally and (b) theoretically. The oscillation in the experimental data is believed to be an artefact of the oscilloscope's input capacitance. The predicted response settles to within 10$\%$ of the final leakage current of the transistor $I_{DS}$ after approximately 2500\,ns from the moment $Q_1$ begins to switch off. }
	\label{Transient test demag}
\end{figure}
\indent Two tests were conducted to analyse the demagnetization circuit's performance, including both transient and steady-state responses. These measurements have to be performed quickly (100s of ns) with very high dynamic range (A~to~pA), hence are difficult to perform simultaneously. Accordingly, the transient response was measured separately using an oscilloscope (Micsig STO1152c), offering the necessary sample rate and current resolution ($\approx 4$\,mA) to observe the avalanche breakdown clamping mechanism and the bulk of the overall response. Secondly, a precision source meter unit (SMU) evaluated the steady-state response, i.e., the leakage current through the drain $I_{DS}$ of $Q_1$. \\
\indent For the transient response test, a 250\,m$\Omega$ shunt resistor was inserted in series with the coil to monitor its current. The oscilloscope was connected to the shunt with a short (15\,cm) coaxial cable to limit its capacitance. A 1\,kHz pulse with a duty cycle of 10\,\% was used to test the demagnetization. This was then compared to simulated data with similar components as exact macro-models were not available. $V_S$ was set to 5.8\,V and the gate driver voltage was set to 12\,V, to mimic typical experimental operating conditions. It can be seen in Fig. \ref{Transient test demag} that the $90\,\%$ to $10\,\%$ experimental ($t_f=215$\,ns) and simulated ($t_f=203$\,ns) fall times are in close agreement. The slight discrepancy most likely arises from the inability to match the exact macro-models, high variation in the breakdown voltage of the TVS diode, and other parasitics present. \\
\indent Next, the leakage current flowing through the transistor's source and drain, $I_{DS}$, was estimated during switch off using a high precision SMU due to the oscilloscope's limited vertical resolution. Only $Q_1$ was tested in this instance. The SMU was configured to source the voltage $V_S$ of $5.8\,$V through $Q_1$. The gate of $Q_1$ was driven by a set of two 9\,V, PP3 batteries connected in series, and then through a potentiometer to set the gate voltage to 12\,V. $Q_1$ was connected to the SMU through a triaxial cable (Balden 9922), to limit any leakage current from the cable assembly or instrument itself \cite{Keithley_LLH2016}. The leakage current of $Q_1$ was found to be lower than 50\,pA, which translates to approximately $135\,$fT based on the theoretically predicted field-to-current ratio of the coils. The response settles to within $10\,\%$ of this steady state value after approximately 2500\,ns from the moment $Q_1$ is switched off. 

\section{Results}
\subsection{FID signal analysis}
\begin{figure}
	\centering
	\includegraphics[scale = 1]{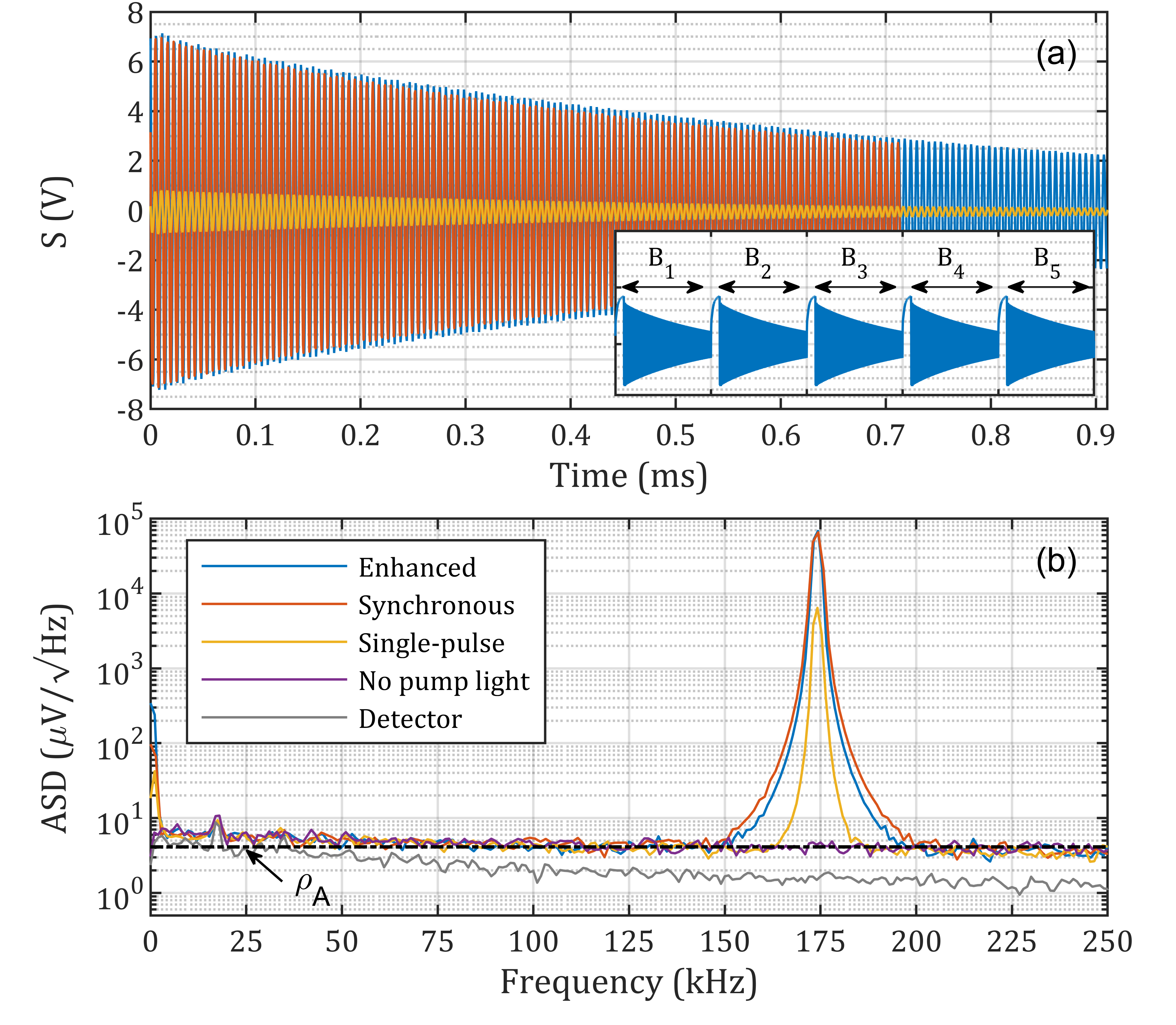}
	\caption{(a) Polarimeter-detected FID traces acquired using different optical pumping strategies including EOP, (blue), synchronous (orange), and SP (yellow). (b) Amplitude-spectral-density (ASD) curves of polarimeter output. Noise spectra were collected for each optical pumping regime, with no pump light applied (purple), and without probe light (grey). $\rho_{A}$ is the average noise density, when no pump light was present, across a $2\,$kHz range centred at the Larmor frequency $\omega_{L} \approx 2\pi \times 175\,$kHz.}
    \label{field-switching comparison fid 50uT}
\end{figure}

\vspace{1pt}
\indent During $T_{r}$, the polarized spins undergo Larmor precession in the presence of $\vec{B_{m}}$. Consequently, the alkali vapor experiences a modulated birefringence, detectable through optical rotation in the linearly polarized probe beam that is monitored by a balanced polarimeter. The optical rotation angle is proportional to the projection of spin polarization along the $x$-axis given by,
\begin{equation}
\label{DS model}
    M_{x}(t) = M_{0}\,\mathrm{sin}(\omega_{L}\,t + \phi_{0})\,e^{-\gamma_{2}t}, 
\end{equation}
 where $M_{0}$ is spin polarisation generated through optical pumping and $\phi_{0}$ is the initial phase. It can be seen that it exhibits a sinusoidal decay where $\omega_{L}$ corresponds to the precession experienced by the Cs atoms, and $\gamma_{2}$ is dependent on the intrinsic properties of the vapor cell and operational systematics. \\
 \indent The photodetector signal is digitized by a data acquisition (DAQ) system based on a Picoscope (model 5444D) operating with 15-bit voltage resolution at a sampling rate of $125\,$Mz. The discretized signal can be modelled as,
\begin{equation}
    S_n = A\,\mathrm{sin}\bigg(\omega_{L}\,n\,\Delta{t} + \phi_{0}\bigg)\,e^{-\gamma_{2}\,n\,\Delta{t}} + \epsilon_n, 
\end{equation}
where $A$ is the FID amplitude, $n$ is the data point of interest, $\Delta{t}$ is the time interval between adjacent samples, and $\epsilon_n$ is the signal noise. Figure \ref{field-switching comparison fid 50uT}(a) depicts examples of FID traces captured by the polarimeter during readout. Each signal was obtained using the optical pumping strategies described previously in a field $B_{y} \approx 50\,\mu\mathrm{T}$. The signal is automatically downsampled by the DAQ device which averages $50$ successive points yielding a final sampling rate of $1/\Delta{t} = 2.5\,$MHz. One can generate a FID signal train by optically pumping the atoms and measuring the subsequent decay of spin polarization over multiple cycles, as seen in the inset of Fig. \ref{field-switching comparison fid 50uT}(a). This method was used throughout these experiments to generate a magnetic field time series at a sampling rate, $f_{d} = 1\,$kHz, resulting in a Nyquist limited bandwidth of $500\,$Hz \cite{hunter2018waveform}. \\
\indent The magnetometer noise budget was assessed by computing the ASD, as shown in Fig. \ref{field-switching comparison fid 50uT}(b) for each optical pumping technique. The ASDs were formulated using Welch's method \cite{welch1967use} by averaging the discrete Fourier transforms (DFTs) calculated from 20 subsequent FID traces over the period, $T_{r}$. A Hanning window was utilized to provide a more accurate determination of the baseline noise levels. Noise spectra were also collected with no pump light applied to the atoms and in the absence of probe light, to localize the dominant noise contributions in the system. These spectra were generated using 20 separate time domain traces as before, although over the full $1\,$ms window set by $f_{d}$ as there is no optical pumping in these cases. \\
\indent Figure \ref{field-switching comparison fid 50uT}(b) shows that there is a close match in the baseline noise levels obtained from the FID spectra and the noise spectrum observed with no optical pumping, i.e., only probe light is present. This is to be expected as the pump beam is mostly extinguished by the AOM during readout, and is further attenuated by the bandpass filter prior to reaching the detector. The spectral peak for the synchronous case is slightly wider due to the shorter measurement window as more time is dedicated to optical pumping. The noise density, $\rho_{A}$, was estimated to be $4\,\mu\mathrm{{V}/\sqrt{Hz}}$, calculated as the average noise density across a $2\,$kHz range centred at the Larmor frequency $\omega_{L} \approx 2\pi \times 175\,$kHz. This noise level dictates the achievable SNR and consequently limits the precision of the Larmor frequency measurement based on the CRLB condition (see Section \ref{sensitivity estimation}). The magnetometer is mostly limited by photon shot-noise of the probe light at this frequency as the technical noise inherent to the detector is less pronounced. This can be attributed to the effective suppression in common-mode noise sources facilitated by balanced photodetection e.g., laser intensity or frequency fluctuations.  \\
\indent The photon shot-noise density can be estimated based on the the detected optical power, $P_{det}$, as,
\begin{equation}
    \rho_{sn} = G\,\sqrt{2\,e\,P_{det}\,\mathcal{R}},
\end{equation}
where $e$ is the electron charge, $\mathcal{R}$ is the detector responsivity, and $G$ is the amplifier transimpedance gain \cite{lucivero2022femtotesla}. With approximately $63\%$ of the probe light reaching the detector, $\rho_{sn}$ was calculated to be $3.7\,\mu$V$/\sqrt{\mathrm{Hz}}$. This is consistent with the noise density, $\rho_{A}$, denoted in Fig. \ref{field-switching comparison fid 50uT}(b) when added in quadrature with the intrinsic noise of the detection system which was measured to be $1.5\,\mu$V$/\sqrt{\mathrm{Hz}}$ at $175\,$kHz.

\subsection{Sensitivity estimation} \label{sensitivity estimation}
\vspace{1pt}
The sensitivity performance of a FID-based magnetometer can be assessed using the CRLB which is a measure of the minimum statistical uncertainty of determining an unbiased estimator from a signal \cite{moschitta2007cramer}. Assuming $\epsilon_n$ is distributed as white Gaussian noise, the CRLB standard deviation for extracting $\omega_{L}$ from a FID trace can be calculated as \cite{yao1995cramer},
\begin{equation}
\label{CRLB}
\sigma_{CR} \geq \frac{\sqrt{12\,C}}{\gamma\,(A/\rho_{A})\,T_{r}^{3/2}}, 
\end{equation}
\noindent where $A$ is the FID amplitude, and $\rho_{A}$ is the noise spectral density at the Larmor frequency. $T_{r}$ is the readout duration, and $N = T_{r}/\Delta{t}$ is the number of samples in the FID trace. $C$ is a correction factor accounting for spin depolarization at a rate, $\gamma_{2}$, and is given by \cite{gemmel2010ultra}, \\
\begin{equation}
\label{Damping factor}
 C = \frac{N}{12} \frac{(1-z^2)^3(1-z^{2N})}{z^2(1-z^{2N})^2 - N^2z^{2N}(1-z^2)^2},
 \end{equation}
\noindent where $z = e^{-\gamma_{2}/N T_{r}}$. The correction factor has a lower bound of unity for an undamped sinusoid. $\sigma_{CR}$ can be converted to a noise density $\rho_{CR} = \sigma_{CR}/\sqrt{f_{d}/2}$ to provide a sensitivity metric in units of $\mathrm{T/\sqrt{Hz}}$. This assumes the magnetometer is white noise limited such that the noise density is flat across all frequencies within the magnetometer bandwidth. \\
\begin{figure}
    \centering
    \includegraphics{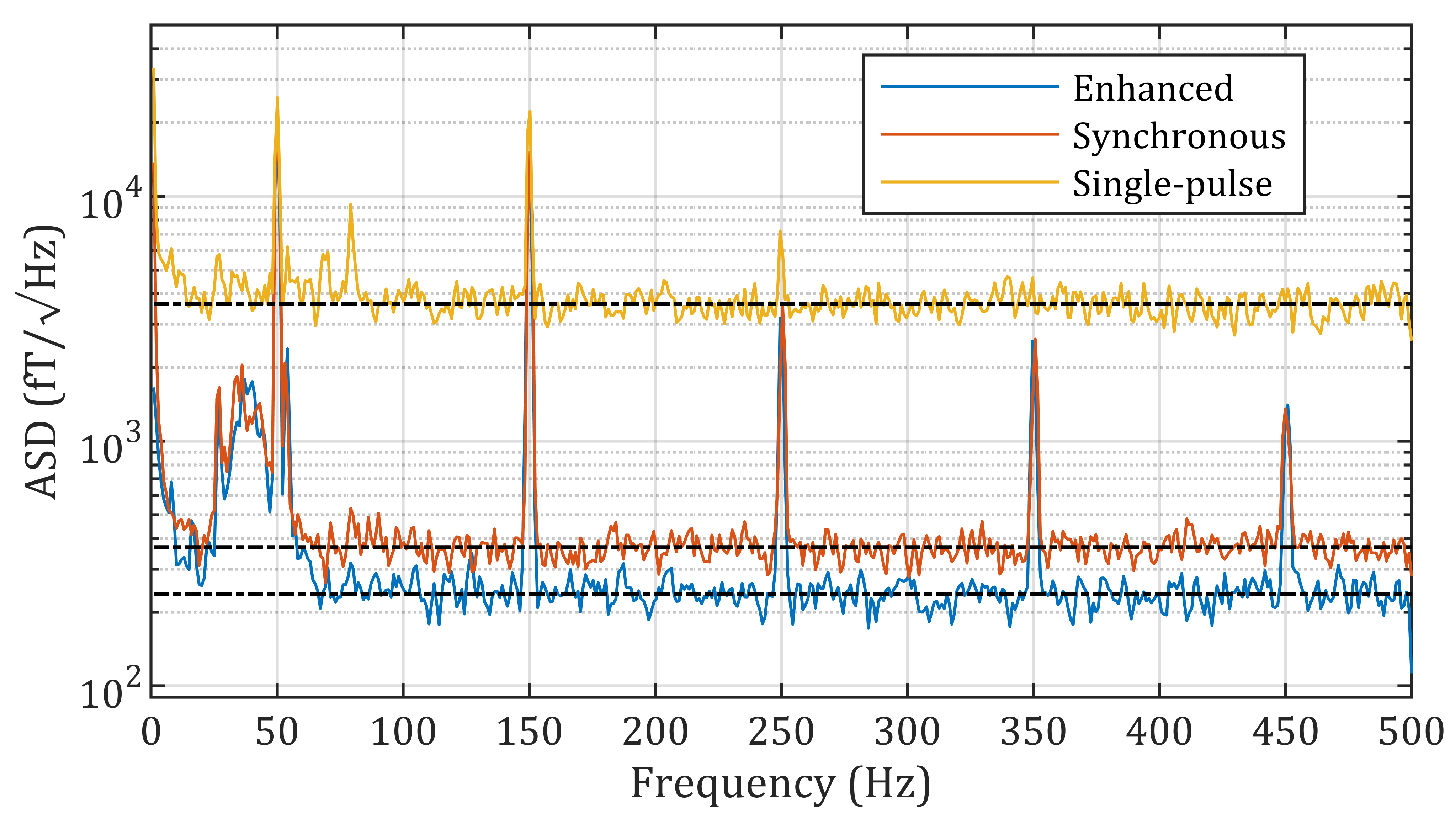}
    \caption{Magnetic field sensitivity spectra acquired using enhanced (blue), synchronous (orange), and single-pulse (yellow) optical pumping. Each ASD was acquired with $B_{y} \approx 50\,\mu$T. The noise floors (dash-dotted lines) were calculated by averaging the spectra over a $70-500\,$Hz frequency range, whilst avoiding technical noise peaks. The corresponding sensitivity estimations, $\rho_{B}$, are listed in Table \ref{table:performance}.}
    \label{sensitivity 50uT}
\end{figure} 
\begin{figure*}
	\centering
	\includegraphics[scale=0.9]{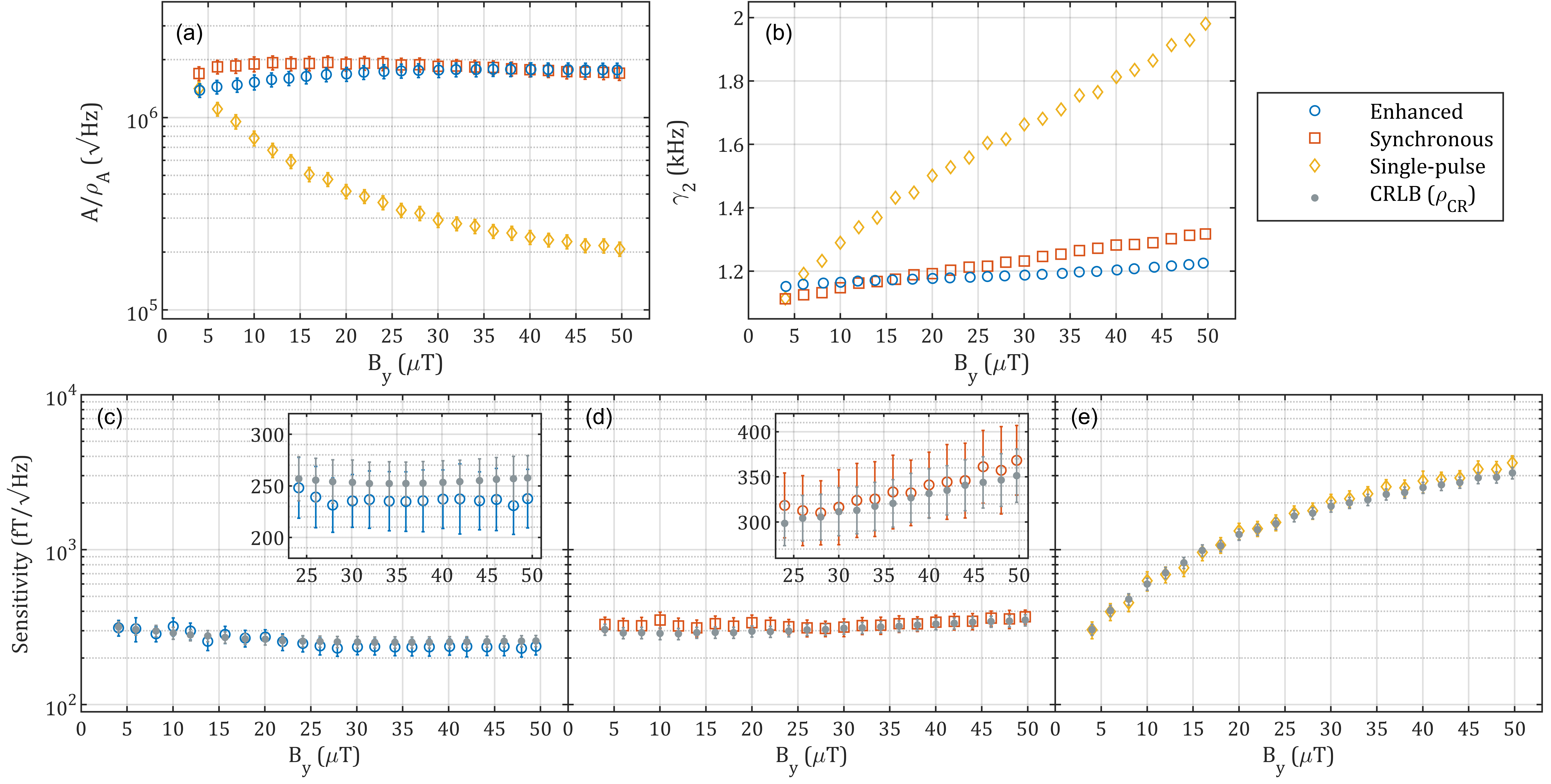}
	\caption{FID magnetometer dynamic range comparison for three optical pumping strategies: EOP (blue), synchronous (orange), and SP (yellow). The bias field, $B_{y}$, was varied between $4-50\,\mu$T. (a) SNR based on the fitted FID amplitude, $A$, and the noise density, $\rho_{A}$, at the Larmor frequency of interest. (b) Transverse relaxation rate, $\gamma_2$, calculated from a single decay period ($1/e$) measured by fitting each FID trace to the model in Eq. \ref{DS model}. The markers are larger than the associated error bars. (c-e) Noise floor, $\rho_B$, estimated from the ASDs (colors), and the associated CRLB noise density, $\rho_{CR}$ (grey).}
	\label{field-switching comparison}
\end{figure*}
\begin{table}[b]
\centering
\small
\begin{tabular}{c c c c c}
 \hline
 \textbf{Parameter} & \textbf{EOP} & \textbf{Synchronous} & \textbf{SP} & \textbf{Units} \\[0.5ex] 
 \hline
 $\mathbf{B_y}$ & 49.5 & \small{49.7} & 49.8 & $\mu$T \\[-0.5ex] 
 $\mathbf{T_p}$ & 88.4 & 286 & 88.4 & $\mu$s \\[-0.5ex]
 $\mathbf{A}$ & 7.29 & \small{7.02} & 0.86 & V \\[-0.5ex] 
 $\boldsymbol{\gamma_{2}}$ & 1.23 & 1.32 & 1.98 & kHz \\[-0.5ex]  
 $\boldsymbol{\rho_{B}}$ & $238 \pm 28.4$ & $368 \pm 38.5$ & $3630 \pm 400$ & fT/$\sqrt{\mathrm{Hz}}$ \\[-0.5ex] 
 $\boldsymbol{\rho_{CR}}$ & $258 \pm 21.7$ & $351 \pm 29.5$ & $3130 \pm 263$ & fT/$\sqrt{\mathrm{Hz}}$ \\[0.5ex] 
 \hline
\end{tabular}
\caption{List of experimental parameters for each implemented optical pumping strategy: EOP, synchronous, and SP. The amplitude, $A$, decay rate, $\gamma_2$, and magnetic field, $B_{y}$, were determined from a damped sinusoidal fit. $\rho_{B}$ and $\rho_{CR}$ are the sensitivity estimations based on the noise floor (see Fig. \ref{sensitivity 50uT}) and CRLB (see Eq. \ref{CRLB}) predicted noise densities, respectively.}
\label{table:performance}
\end{table}
\indent An alternative definition of sensitivity is based on the sensor noise floor, $\rho_B$, which can be obtained by monitoring magnetic field fluctuations recorded by the OPM over a set time interval. A magnetic field time series can be produced by extracting $\omega_{L}$ from consecutive FID traces in a signal train over multiple cycles as illustrated in the inset of Fig. \ref{CRLB}(a). Subsequently, the ASD in fT$/\sqrt{\mathrm{Hz}}$ can be generated by applying the DFT to the time series data, as performed in previous instances. Figure \ref{sensitivity 50uT} shows a set of magnetic field sensitivity spectra gathered using this approach for each of the aforementioned optical pumping strategies in a bias field, $B_y \approx 50\,\mu$T. The ASD curves in Fig. \ref{sensitivity 50uT} were computed in a similar manner to before by averaging 20 non-overlapping $1\,$s time segments using Welch's method. The technical noise peaks observed between the frequencies $20-60\,$Hz are related to environmental magnetic noise penetrating the three-layer $\mu$-metal shield encapsulating the vapor cell. Additional noise peaks at $50\,$Hz, and associated harmonics, originate from the current supply driving the coil producing the bias field. The noise floor, $\rho_B$, and associated uncertainty were determined by calculating the average and standard deviation of the noise density over a $70-500\,$Hz frequency range, ignoring technical noise peaks. \\
\indent The raw FID traces shown in Fig. \ref{CRLB}(a) were processed post-acquisition by fitting the data to the model given in Eq. \ref{DS model}, providing the amplitude $A$, and decay rate, $\gamma_{2}$, measurements required to calculate $\rho_{CR}$. The corresponding sensitivity estimations are listed in Table \ref{table:performance} for each optical pumping technique along with other relevant experimental parameters. The error in $\rho_{CR}$ is mostly attributed to the uncertainty in the estimation of $\rho_{A}$. Evidently, these sensitivity estimations are closely correlated with $\rho_{B}$ for each optical pumping strategy. This clearly shows that external magnetic noise contributions, e.g., produced by stray currents in the heater, are not a limiting factor. Considering the white noise assumption in determining the CRLB, this further validates that the magnetometer is predominantly limited by photon shot-noise.    \\

\subsection{Dynamic range characterization}
\vspace{1pt}
\indent Dynamic range represents the limits in magnetic field that a magnetometer can reliably operate within and plays a crucial role in finite-field sensing. Some OPMs, e.g., SERF systems, can only function close to zero-field requiring well-conditioned magnetic field environments. In contrast, the FID-based approach enables operation over a wide range of bias fields as observed in Fig. \ref{field-switching comparison}. This conveys the sensors performance across the range $B_{y} \approx 4-50\,\mu$T and provides a comparison of each optical pumping method discussed in this work. \\
\indent Clearly, SP optical pumping performs relatively poorly compared to synchronous modulation and EOP. This is particularly evident in Fig. \ref{field-switching comparison}(a) which shows a significant degradation in SNR as $B_y$ is raised. This is directly related to a reduction in $A$ as $\rho_{A}$ is consistent for each pumping scheme as seen in Fig. \ref{CRLB}(b). As mentioned previously, this degradation arises from transverse fields, e.g., $\vec{B_m}$, applying a torque on the spin polarization during optical pumping, and is well described by the Bloch equation formalism \cite{grujic2015sensitive}. Figure \ref{field-switching comparison}(b) shows the additional impact this has on $\gamma_2$ which experiences a sharp rise as $B_{y}$ is raised. This is to be expected as the effectiveness of spin-exchange suppression diminishes when the spins are prepared into a less polarized state.    \\
\indent A more gradual reduction in $A$ is observed at elevated bias fields when employing either synchronous modulation or EOP, although this is not immediately apparent from Fig. \ref{field-switching comparison}(a) as the SNR stays relatively consistent. This is because $\rho_{A}$ also reduces as $B_{y}$ is increased; a consequence of the technical noise present in the detection system being less prominent at higher Larmor frequencies (see Fig. \ref{CRLB}(b)). The $1/f$ noise dependence of $\rho_{A}$ also explains the drop in SNR at lesser $B_y$ values. Slightly higher signal amplitudes were achieved with synchronous modulation at low bias fields owing to the longer pumping period used. \\
\indent It is anticipated that much of the loss in amplitude observed for both EOP and synchronous modulation at larger bias fields will be attributed to nonlinear Zeeman (NLZ) splitting. This effect simultaneously broadens and distorts the magnetic resonance, and is more prevalent at stronger field magnitudes \cite{bao2018suppression}. In the case of EOP, this broadening mechanism will be the main contributor to the $\approx 80\,$Hz deviation in $\gamma_2$ shown in Fig. \ref{field-switching comparison}(b), over the range of fields tested. Further investigation is necessary to fully quantify the effects NLZ has on the spin dynamics. Broadening due to magnetic field gradients should also be considered, although the deviation in magnetic field expected across the beam width based on the coil geometry is $\Delta{B} \approx 5.5\,$nT. Gradient broadening is determined by the spread of precession frequencies throughout the cell, i.e., $\gamma_{gr} \sim \gamma\,\Delta{B}$ \cite{seltzer2008developments}, which is around $19\,$Hz in this case.   \\
\indent The integrated area of the magnetic resonance, in the frequency domain, is a reflection of the spin polarization gained through optical pumping. Thus, assuming this remains constant, one would expect a lower signal amplitude given the additional broadening that is induced by the NLZ effect. EOP demonstrated consistent spin polarization buildup across the full range of bias fields tested. This is not surprising given that the field strength of $\vec{B_{p}}$ is at least two orders of magnitude higher than $B_{y}$, thus the spin polarization will no longer experience a significant torque during pumping. In contrast, the steady-state spin polarization achieved with synchronous modulation reduced as a function of $B_{y}$. NLZ splitting inevitably influences the optical pumping dynamics in this case, as the atoms become more difficult to resonantly address due to the nonlinearity in the magnetic sublevel structure. Naturally, this will result in a steeper amplitude reduction with respect to $B_{y}$ compared to EOP which does not rely on a resonant driving field. Furthermore, this accounts for the sharper rise in $\gamma_2$ for synchronous pumping as spin-exchange suppression becomes less effective.  \\
\indent Figures \ref{field-switching comparison}(c-e) provide a comparison in sensitivity performance for each of the optical pumping schemes investigated. Two approaches to sensitivity estimation were used in accordance with that described in Section \ref{sensitivity estimation}. The first metric is the CRLB noise density, $\rho_{CR}$, denoted by the gray data points. The second method estimates the noise floor, $\rho_{B}$, of the ASD computed from the magnetic field recordings, represented as colored markers. It can be readily seen that the magnetometer noise floor reaches the CRLB limit across the full range of bias fields for each pumping technique. This validates that the heating strategy used in EOP is not lifting the sensor noise level, which is to be expected since the coil is rapidly demagnetized over a short period of $200\,$ns. It also verifies that the magnetic field fluctuations produced by the bias field are well below the noise floor of the sensor. The sensitivity dependence matches well with expectations in accordance with Eq. \ref{CRLB} when considering the SNR and $\gamma_2$ achieved in each case. The best sensitivity was obtained using EOP as seen in the inset of Fig. \ref{field-switching comparison}(c), due to the improved signal amplitudes and relaxation rates achieved with more efficient optical pumping, especially at large magnetic field strengths.

\subsection{Accuracy Considerations}
\begin{figure}
    \centering
    \includegraphics[scale = 1]{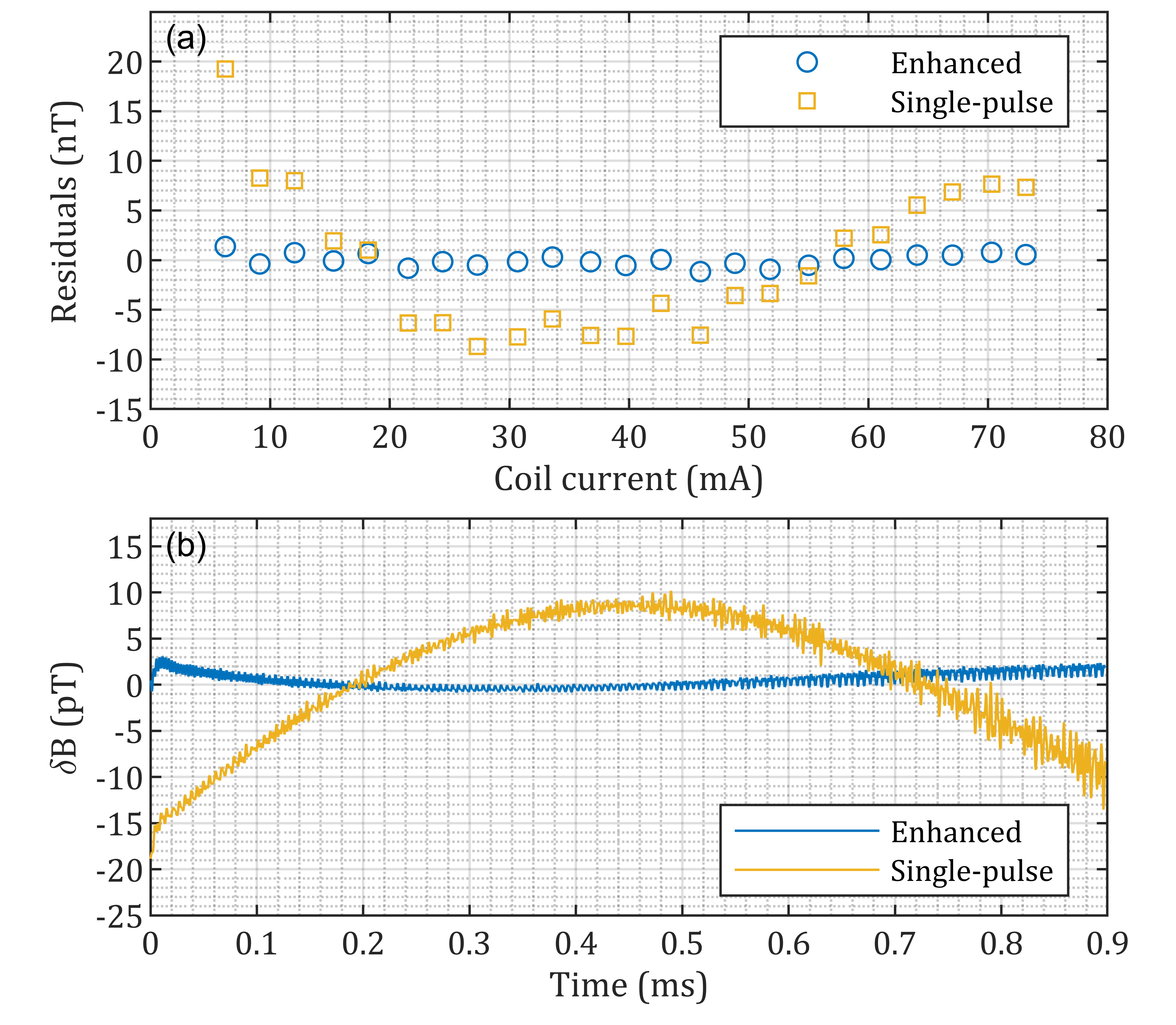}
    \caption{(a) Linear fit residuals of magnetic field data recorded at various bias coil currents, using EOP (blue squares) and SP modulation (yellow circles). (b) Residual variation in the instantaneous magnetic field recorded over a FID cycle using EOP and SP modulation as noted in the legend. The data was processed using a Hilbert transform to calculate the spin precession phase at each DAQ clock cycle, and hence the instantaneous precession frequency. The plot above shows the residual variation from a linear phase dependence, and is an average of 50 consecutive FID pulses.}
    \label{Accuracy}
\end{figure}

\vspace{1pt}
\indent FID-based sensors have a distinct advantage in accuracy compared to other OPM configurations as they are inherently self-calibrating and do not suffer from light shifts caused by intense optical pumping. Despite this, systematics are still present when operating in geophysical magnetic fields due to magnetic resonance asymmetries \cite{zhang2023heading}. For example, both hyperfine ground states possess slightly different gyromagnetic factors, and subsequently, their precession frequencies diverge depending on the magnetic field strength. Moreover, the magnetic resonance can be distorted owing to NLZ splitting of a single hyperfine manifold \cite{bao2018suppression}. In both cases, the induced systematics depend on the atomic distribution among the Zeeman sublevels of both hyperfine levels, which is sensitive to both spin preparation and subsequent relaxation. \\
\indent The linearity of the magnetometer response with magnetic field was characterized for the EOP and SP modulation schemes. The average magnetic field was monitored over $1\,$s measurement intervals at various coil supply currents as seen in Fig.~\ref{Accuracy}(a). This shows the residuals obtained from a linear fit to the recorded magnetic field data. The magnetometer responds relatively linearly with bias field in the case of EOP as the residuals fluctuate around zero with no perceivable trends. EOP reliably generates the same polarization state independent of the applied field strength. Consequently, the atomic population mainly resides in the $F = 4$ ground state such that the spins precess at a single predominant frequency, minimising systematic effects. The residuals for SP modulation convey a distinct quadratic behaviour, as a result of inconsistency in the atomic population distribution achieved through optical pumping in various bias field conditions. In this case, the Larmor frequency estimation will be weighted according to the atomic population occupying each hyperfine state. \\
\indent In the EOP scheme, rapid coil demagnetization is crucial in preventing spin precession readout from being affected by the magnetic field pulse. Previously, this was verified electronically by measuring the transient current response, which demonstrated $\sim 200\,$ns fall time and eventual decay to around $135\,$fT due to the leakage current of the MOSFET. In order to observe the transient magnetic response from the magnetometer, an alternative signal processing strategy was devised based on a Hilbert Transform. This vastly extends the magnetometer bandwidth enabling resolution of frequencies up to the Nyquist limit dictated by the sampling rate of the DAQ system \cite{wilson2020wide}. Generation of the full signal phasor using a dual matched finite-impulse response filter to perform a Hilbert transform allows estimation of the signal phase at each DAQ clock cycle, and hence the instantaneous precession frequency \cite{ingleby2022digital}. Consistent variation over time is observed using the SP scheme, and to a lesser extent, with the EOP scheme, and the trend can be clarified by averaging over 50 consecutive FID pulses (see Fig. \ref{Accuracy}(b)). 
\\
\indent Figure \ref{Accuracy}(b) shows the residual variation in instantaneous magnetic field over the course of a single FID cycle in a bias field $B_{y} \approx 50\,\mu$T. The observations for SP optical pumping suggest that the observed systematic variation of precession frequency during each FID pulse is not a consequence of demagnetization, as no current was flowing through the PCB coils using this technique. Instead, the decay of spin polarisation during the FID pulse, combined with the dependence of NLZ-induced heading error on this polarisation, is likely to be responsible for this behaviour. Ground-state spin relaxation creates a time-dependent spin distribution, which varies consistently during each FID pulse. At high bias fields, the splitting between the Zeeman sublevels is no longer linear, creating a dependence of the measured Larmor frequency on the magnetic sublevel populations, which vary during the FID pulse. One would expect the trend to be more distinctive for the SP case as the influence of spin-exchange processes results in more complex evolution of the ground-state spin distriubutions. Significantly, a lower systematic shift is observed in the EOP case, confirming that the pre-polarising coil is sufficiently de-energised at the measurement onset.      

\section{Conclusion and Outlook}
\vspace{1pt}
In conclusion, a more efficient spin preparation scheme for FID atomic magnetometers was demonstrated, by applying mT-level magnetic field pulses synchronized with the amplitude modulation of the pump light. A PCB coil pair encapsulates the miniaturized Cs vapor cell to produce this strong magnetic field along the optical pumping axis, whilst simultaneously resistively heating the vapor to the optimal atomic density for maximum sensitivity performance. No magnetic noise is produced as the coils are only active when optically pumping the atoms, and are rapidly demagnetized to near zero prior to spin readout. Tests of the demagnetization circuit's transient response yielded a fall time of $213\,$ns, closely matching theoretical predictions. Furthermore, a digital Hilbert transform applied to individual FID traces revealed no signature of the demagnetization process; this is noteworthy given that accuracy is one of the key benefits of the FID modality. The spin polarization prepared through EOP is consistent over a wide range of bias fields which aids in maintaining the sensor accuracy. Systematics in the Larmor frequency measurement induced by heading errors remain stable under various bias fields. Heading error can be corrected analytically based on the degree spin polarization \cite{lee2021heading}; however, this compensation becomes more straightforward under consistent optical pumping conditions. \\
\indent An optimal magnetic sensitivity of $238\,\mathrm{fT/\sqrt{Hz}}$ was achieved in a bias field of $50\,\mu$T using EOP providing an improvement over existing optical pumping schemes. The sensor noise floor closely matched CRLB noise density predictions indicating that magnetic noise, e.g., arising from electrical heating or the bias field, is not a limiting factor. Therefore, sensitivity improvement can only be achieved by increasing the signal amplitude, as the magnetometer is mostly photon shot-noise limited especially at higher Larmor frequencies, or extending the spin coherence time. Optical rotation experienced by the probe beam could be doubled by placing a reflector after the vapor cell in a double-pass geometry. A compact device would also benefit from this geometry due to the reduced standoff distance between the vapor cell and the signal source of interest. Recent novel fabrications techniques \cite{dyer2022micro} make it feasible to manufacture thicker MEMS cells that exhibit higher signal amplitudes and longer spin relaxation times. Furthermore, the customizable cell geometries improve the available optical access that can be interrogated. One could also lower the noise density, $\rho_{A}$, slightly by implementing a faster low-noise DAQ system with higher bit-resolution, e.g., using field programmable gate arrays (FPGAs). \\
\indent The EOP strategy is ideally suited to sensor commercialization as it provides a scalable solution that limits the hardware and software requirements for robust operation. For example, sensors based on resonantly driven spin precession require electronic feedback loops to maintain optimal performance which increases complexity. Furthermore, more elaborate cell heating methods are required to prevent stray magnetic fields from raising the sensor noise floor. The magnetometer's topology could be made more compact and portable by replacing the pump and probe lasers with a vertical-cavity surface-emitting laser (VCSEL). The pump light would be frequency modulated in this case, as opposed to utilizing amplitude modulation with an AOM \cite{hunter2018free}. The power consumption of the pre-polarizing field and cell heating in the current configuration is approximately 0.8 W. This sort of power can be easily accessible with USB-C \cite{usb-c2022}. The power draw can be further optimized by the modification to coil geometry, balancing current draw and field induced. One could envision a high-performance finite-field sensor with minimal optical components including: at least one VCSEL for pumping and probing, a quarter-wave plate, MEMS vapor cell, PCB coils for EOP and heating, and a balanced detector.

\begin{backmatter}
\bmsection{Funding} Innovate UK (ISCF-42186).

\bmsection{Acknowledgments} AM was supported by a Ph.D. studentship from the Defence Science and Technology Laboratory (Dstl).

\bmsection{Disclosures} The authors declare that there are no conflicts of interest related to this article.

\bmsection{Data availability} Data underlying the results presented in this manuscript are available in Ref. \cite{hunter2023free}.
\end{backmatter}

\smallskip

% Bibliography
\bibliography{references}

\providecommand{\noopsort}[1]{}\providecommand{\singleletter}[1]{#1}%
\begin{thebibliography}{10}
\newcommand{\enquote}[1]{``#1''}

\bibitem{cohen1970diverses}
C.~Cohen-Tannoudji, J.~Dupont-Roc, S.~Haroche, and F.~Lalo{\"e},
  \enquote{Diverses r{\'e}sonances de croisement de niveaux sur des atomes
  pomp{\'e}s optiquement en champ nul ii. applications {\`a} la mesure de
  champs faibles,} {\protect\JournalTitle{Revue de Physique Appliqu{\'e}e}}
  \textbf{5}, 102--108 (1970).

\bibitem{castagna2011measurement}
N.~Castagna and A.~Weis, \enquote{Measurement of longitudinal and transverse
  spin relaxation rates using the ground-state {H}anle effect,}
  {\protect\JournalTitle{Physical Review A}} \textbf{84}, 053421 (2011).

\bibitem{jimenez2014optically}
R.~Jimenez-Martinez, S.~Knappe, and J.~Kitching, \enquote{An optically
  modulated zero-field atomic magnetometer with suppressed spin-exchange
  broadening,} {\protect\JournalTitle{Review of Scientific Instruments}}
  \textbf{85}, 045124 (2014).

\bibitem{allred2002high}
J.~Allred, R.~Lyman, T.~Kornack, and M.~V. Romalis, \enquote{High-sensitivity
  atomic magnetometer unaffected by spin-exchange relaxation,}
  {\protect\JournalTitle{Physical Review Letters}} \textbf{89}, 130801 (2002).

\bibitem{shah2009spin}
V.~Shah and M.~V. Romalis, \enquote{Spin-exchange relaxation-free magnetometry
  using elliptically polarized light,} {\protect\JournalTitle{Physical Review
  A}} \textbf{80}, 013416 (2009).

\bibitem{li2018serf}
J.~Li, W.~Quan, B.~Zhou, Z.~Wang, J.~Lu, Z.~Hu, G.~Liu, and J.~Fang,
  \enquote{Serf atomic magnetometer--recent advances and applications: A
  review,} {\protect\JournalTitle{IEEE Sensors Journal}} \textbf{18},
  8198--8207 (2018).

\bibitem{dang2010ultrahigh}
H.~Dang, A.~C. Maloof, and M.~V. Romalis, \enquote{Ultrahigh sensitivity
  magnetic field and magnetization measurements with an atomic magnetometer,}
  {\protect\JournalTitle{Applied Physics Letters}} \textbf{97}, 151110 (2010).

\bibitem{boto2017new}
E.~Boto, S.~S. Meyer, V.~Shah, O.~Alem, S.~Knappe, P.~Kruger, T.~M. Fromhold,
  M.~Lim, P.~M. Glover, P.~G. Morris, R.~Bowtell, G.~R. Barnes, and M.~J.
  Brookes, \enquote{A new generation of magnetoencephalography: Room
  temperature measurements using optically-pumped magnetometers,}
  {\protect\JournalTitle{NeuroImage}} \textbf{149}, 404--414 (2017).

\bibitem{osborne2018fully}
J.~Osborne, J.~Orton, O.~Alem, and V.~Shah, \enquote{Fully integrated
  standalone zero field optically pumped magnetometer for biomagnetism,} in
  \emph{Steep dispersion engineering and opto-atomic precision metrology XI,}
  vol. 10548 (SPIE, 2018), pp. 89--95.

\bibitem{boto2018moving}
E.~Boto, N.~Holmes, J.~Leggett, G.~Roberts, V.~Shah, S.~S. Meyer, L.~D.
  Mu{\~n}oz, K.~J. Mullinger, T.~M. Tierney, S.~Bestmann, G.~R. Barnes,
  R.~Bowtell, and M.~J. Brookes, \enquote{Moving magnetoencephalography towards
  real-world applications with a wearable system,}
  {\protect\JournalTitle{Nature}} \textbf{555}, 657--661 (2018).

\bibitem{holmes2019balanced}
N.~Holmes, T.~M. Tierney, J.~Leggett, E.~Boto, S.~Mellor, G.~Roberts, R.~M.
  Hill, V.~Shah, G.~R. Barnes, M.~J. Brookes, and R.~Bowtell,
  \enquote{Balanced, bi-planar magnetic field and field gradient coils for
  field compensation in wearable magnetoencephalography,}
  {\protect\JournalTitle{Scientific Reports}} \textbf{9}, 14196 (2019).

\bibitem{gerginov2017pulsed}
V.~Gerginov, S.~Krzyzewski, and S.~Knappe, \enquote{Pulsed operation of a
  miniature scalar optically pumped magnetometer,} {\protect\JournalTitle{JOSA
  B}} \textbf{34}, 1429--1434 (2017).

\bibitem{guo2019compact}
Y.~Guo, S.~Wan, X.~Sun, and J.~Qin, \enquote{Compact, high-sensitivity atomic
  magnetometer utilizing the light-narrowing effect and in-phase excitation,}
  {\protect\JournalTitle{Applied Optics}} \textbf{58}, 734--738 (2019).

\bibitem{lucivero2022femtotesla}
V.~Lucivero, W.~Lee, T.~Kornack, M.~Limes, E.~Foley, and M.~Romalis,
  \enquote{Femtotesla nearly-quantum-noise-limited pulsed gradiometer at
  earth-scale fields,} {\protect\JournalTitle{Physical Review Applied}}
  \textbf{18}, L021001 (2022).

\bibitem{beggan2018observation}
C.~Beggan and M.~Musur, \enquote{Observation of ionospheric {A}lfv{\'e}n
  resonances at 1--30 {H}z and their superposition with the {S}chumann
  resonances,} {\protect\JournalTitle{Journal of Geophysical Research: Space
  Physics}} \textbf{123}, 4202--4214 (2018).

\bibitem{korth2016miniature}
H.~Korth, K.~Strohbehn, F.~Tejada, A.~G. Andreou, J.~Kitching, S.~Knappe, S.~J.
  Lehtonen, S.~M. London, and M.~Kafel, \enquote{Miniature atomic scalar
  magnetometer for space based on the rubidium isotope $^{87}${R}b,}
  {\protect\JournalTitle{Journal of Geophysical Research: Space Physics}}
  \textbf{121}, 7870--7880 (2016).

\bibitem{canciani2021magnetic}
A.~J. Canciani, \enquote{Magnetic navigation on an {F}-16 aircraft using online
  calibration,} {\protect\JournalTitle{IEEE Transactions on Aerospace and
  Electronic Systems}} \textbf{58}, 420--434 (2021).

\bibitem{limes2020portable}
M.~Limes, E.~Foley, T.~Kornack, S.~Caliga, S.~McBride, A.~Braun, W.~Lee,
  V.~Lucivero, and M.~Romalis, \enquote{Portable magnetometry for detection of
  biomagnetism in ambient environments,} {\protect\JournalTitle{Physical Review
  Applied}} \textbf{14}, 011002 (2020).

\bibitem{appelt1999light}
S.~Appelt, A.~B.-A. Baranga, A.~Young, and W.~Happer, \enquote{Light narrowing
  of rubidium magnetic-resonance lines in high-pressure optical-pumping cells,}
  {\protect\JournalTitle{Physical Review A}} \textbf{59}, 2078 (1999).

\bibitem{scholtes2011light}
T.~Scholtes, V.~Schultze, R.~IJsselsteijn, S.~Woetzel, and H.-G. Meyer,
  \enquote{Light-narrowed optically pumped {M}x magnetometer with a
  miniaturized {C}s cell,} {\protect\JournalTitle{Physical Review A}}
  \textbf{84}, 043416 (2011).

\bibitem{grujic2015sensitive}
Z.~D. Gruji{\'c}, P.~A. Koss, G.~Bison, and A.~Weis, \enquote{A sensitive and
  accurate atomic magnetometer based on free spin precession,}
  {\protect\JournalTitle{The European Physical Journal D}} \textbf{69}, 1--10
  (2015).

\bibitem{hunter2018waveform}
D.~Hunter, R.~Jim{\'e}nez-Mart{\'\i}nez, J.~Herbsommer, S.~Ramaswamy, W.~Li,
  and E.~Riis, \enquote{Waveform reconstruction with a {C}s based
  free-induction-decay magnetometer,} {\protect\JournalTitle{Optics express}}
  \textbf{26}, 30523--30531 (2018).

\bibitem{hunter2022accurate}
D.~Hunter, T.~E. Dyer, and E.~Riis, \enquote{Accurate optically pumped
  magnetometer based on {R}amsey-style interrogation,}
  {\protect\JournalTitle{Optics Letters}} \textbf{47}, 1230--1233 (2022).

\bibitem{wilson2020wide}
N.~Wilson, C.~Perrella, R.~Anderson, A.~Luiten, and P.~Light,
  \enquote{Wide-bandwidth atomic magnetometry via instantaneous-phase
  retrieval,} {\protect\JournalTitle{Physical Review Research}} \textbf{2},
  013213 (2020).

\bibitem{ingleby2022digital}
S.~Ingleby, P.~Griffin, T.~Dyer, M.~Mrozowski, and E.~Riis, \enquote{A digital
  alkali spin maser,} {\protect\JournalTitle{Scientific Reports}} \textbf{12},
  1--7 (2022).

\bibitem{hunter2018free}
D.~Hunter, S.~Piccolomo, J.~Pritchard, N.~Brockie, T.~Dyer, and E.~Riis,
  \enquote{Free-induction-decay magnetometer based on a microfabricated {C}s
  vapor cell,} {\protect\JournalTitle{Physical Review Applied}} \textbf{10},
  014002 (2018).

\bibitem{gerginov2020scalar}
V.~Gerginov, M.~Pomponio, and S.~Knappe, \enquote{Scalar magnetometry below 100
  {fT/Hz$^{1/2}$} in a microfabricated cell,} {\protect\JournalTitle{IEEE
  Sensors Journal}} \textbf{20}, 12684--12690 (2020).

\bibitem{jimenez2009sensitivity}
R.~Jim{\'e}nez-Mart{\'\i}nez, W.~C. Griffith, Y.-J. Wang, S.~Knappe,
  J.~Kitching, K.~Smith, and M.~D. Prouty, \enquote{Sensitivity comparison of
  {M}x and frequency-modulated bell--bloom {C}s magnetometers in a
  microfabricated cell,} {\protect\JournalTitle{IEEE Transactions on
  Instrumentation and Measurement}} \textbf{59}, 372--378 (2009).

\bibitem{tayler2022miniature}
M.~C. Tayler, K.~Mouloudakis, R.~Zetter, D.~Hunter, V.~G. Lucivero,
  S.~Bodenstedt, L.~Parkkonen, and M.~W. Mitchell, \enquote{Miniature biplanar
  coils for alkali-metal-vapor magnetometry,} {\protect\JournalTitle{Physical
  Review Applied}} \textbf{18}, 014036 (2022).

\bibitem{sheng2017microfabricated}
D.~Sheng, A.~R. Perry, S.~P. Krzyzewski, S.~Geller, J.~Kitching, and S.~Knappe,
  \enquote{A microfabricated optically-pumped magnetic gradiometer,}
  {\protect\JournalTitle{Applied Physics Letters}} \textbf{110}, 031106 (2017).

\bibitem{kominis2003subfemtotesla}
I.~Kominis, T.~Kornack, J.~Allred, and M.~V. Romalis, \enquote{A subfemtotesla
  multichannel atomic magnetometer,} {\protect\JournalTitle{Nature}}
  \textbf{422}, 596--599 (2003).

\bibitem{dyer2022micro}
S.~Dyer, P.~Griffin, A.~Arnold, F.~Mirando, D.~Burt, E.~Riis, and
  J.~McGilligan, \enquote{Micro-machined deep silicon atomic vapor cells,}
  {\protect\JournalTitle{Journal of Applied Physics}} \textbf{132}, 134401
  (2022).

\bibitem{schultze2015improving}
V.~Schultze, T.~Scholtes, R.~IJsselsteijn, and H.-G. Meyer, \enquote{Improving
  the sensitivity of optically pumped magnetometers by hyperfine repumping,}
  {\protect\JournalTitle{JOSA B}} \textbf{32}, 730--736 (2015).

\bibitem{mrozowski2023}
M.~S. Mrozowski, I.~C. Chalmers, S.~J. Ingleby, P.~F. Griffin, and E.~Riis,
  \enquote{Ultra-low noise, bi-polar, programmable current sources,}
  {\protect\JournalTitle{Review of Scientific Instruments}} \textbf{94}, 014701
  (2023).

\bibitem{PhysRev1954avalanche}
K.~G. McKay, \enquote{Avalanche breakdown in silicon,}
  {\protect\JournalTitle{Phys. Rev.}} \textbf{94}, 877--884 (1954).

\bibitem{microsemi_134}
K.~Walters, \enquote{Zeners and transient voltage suppressors: Can either
  device be used for the same applications?} Tech. Rep. 134, Microsemi (now
  part of Microchip), One Enterprise, Aliso Viejo, CA 92656 USA (2018).

\bibitem{Keithley_LLH2016}
Keithley Instruments, LLC (now part of Tektronix, Inc), Solon, OH, USA,
  \emph{Low Level Measurements Handbook}, 7th ed. (2016).

\bibitem{welch1967use}
P.~Welch, \enquote{The use of fast fourier transform for the estimation of
  power spectra: a method based on time averaging over short, modified
  periodograms,} {\protect\JournalTitle{IEEE Transactions on audio and
  electroacoustics}} \textbf{15}, 70--73 (1967).

\bibitem{moschitta2007cramer}
A.~Moschitta and P.~Carbone, \enquote{Cram{\'e}r--{R}ao lower bound for
  parametric estimation of quantized sinewaves,} {\protect\JournalTitle{IEEE
  Transactions on Instrumentation and Measurement}} \textbf{56}, 975--982
  (2007).

\bibitem{yao1995cramer}
Y.-X. Yao and S.~M. Pandit, \enquote{Cram{\'e}r-{R}ao lower bounds for a damped
  sinusoidal process,} {\protect\JournalTitle{IEEE Transactions on Signal
  Processing}} \textbf{43}, 878--885 (1995).

\bibitem{gemmel2010ultra}
C.~Gemmel, W.~Heil, S.~Karpuk, K.~Lenz, C.~Ludwig, Y.~Sobolev, K.~Tullney,
  M.~Burghoff, W.~Kilian, S.~Knappe-Gr{\"u}neberg, W.~Mueller, A.~Schnabel,
  F.~Seifert, L.~Trahms, and S.~Baessler, \enquote{Ultra-sensitive magnetometry
  based on free precession of nuclear spins,} {\protect\JournalTitle{The
  European Physical Journal D}} \textbf{57}, 303--320 (2010).

\bibitem{bao2018suppression}
G.~Bao, A.~Wickenbrock, S.~Rochester, W.~Zhang, and D.~Budker,
  \enquote{Suppression of the nonlinear zeeman effect and heading error in
  earth-field-range alkali-vapor magnetometers,}
  {\protect\JournalTitle{Physical Review Letters}} \textbf{120}, 033202 (2018).

\bibitem{seltzer2008developments}
S.~J. Seltzer, \emph{Developments in alkali-metal atomic magnetometry}
  (Princeton University, 2008).

\bibitem{zhang2023heading}
R.~Zhang, D.~Kanta, A.~Wickenbrock, H.~Guo, and D.~Budker,
  \enquote{Heading-error-free optical atomic magnetometry in the earth-field
  range,} {\protect\JournalTitle{Physical Review Letters}} \textbf{130}, 153601
  (2023).

\bibitem{lee2021heading}
W.~Lee, V.~Lucivero, M.~Romalis, M.~Limes, E.~Foley, and T.~Kornack,
  \enquote{Heading errors in all-optical alkali-metal-vapor magnetometers in
  geomagnetic fields,} {\protect\JournalTitle{Physical Review A}} \textbf{103},
  063103 (2021).

\bibitem{usb-c2022}
USB 3.0 Promoter Group, \emph{{Universal Serial Bus Type-C Cable and Connector
  Specification}} (2022). Release 2.2.

\bibitem{hunter2023free}
D.~Hunter, M.~S. Mrozowski, A.~McWilliam, S.~J. Ingleby, T.~E. Dyer, P.~F.
  Griffin, and E.~Riis, Data for: "Optical pumping enhancement of a
  free-induction-decay magnetometer", University of Strathclyde,
  https://doi.org/10.15129/49adaba8-c05c-4e18-beca-18f052123984 (2023).

\end{thebibliography}

% Full bibliography added automatically for Optics Letters submissions; the following line will simply be ignored if submitting to other journals.
% Note that this extra page will not count against page length
\bibliographyfullrefs{references}

\end{document}